\newcommand{\bn}{\langle N \rangle}

\documentclass[%
 aip,
pof, amsmath,amssymb,
preprint,%
]{revtex4-1}

\usepackage{graphicx}
\usepackage{bm, ulem}
\usepackage{color}
\definecolor{orange}{rgb}{1,0.3,0.2}
\graphicspath{{figures/}}
\definecolor{vert}{rgb}{0,0.6,0}

\begin{document}


\title{Velocity fluctuations and population distribution in clusters of settling particles at low Reynolds number}

\author{A. Boschan}
\affiliation{Grupo de
Medios Porosos, Departamento de F\'\i sica, Facultad de Ingenier\'\i a,
Universidad de Buenos Aires, Paseo Col\'on 850, 1063 Buenos-Aires, Argentina. }

\author{B.L. Ocampo}
\affiliation{Grupo de
Medios Porosos, Departamento de F\'\i sica, Facultad de Ingenier\'\i a,
Universidad de Buenos Aires, Paseo Col\'on 850, 1063 Buenos-Aires, Argentina. }

\author{M. Annichini}
\affiliation{Grupo de
Medios Porosos, Departamento de F\'\i sica, Facultad de Ingenier\'\i a,
Universidad de Buenos Aires, Paseo Col\'on 850, 1063 Buenos-Aires, Argentina. }

\author{G. Gauthier}
\affiliation{Laboratoire FAST, Univ. Paris-Sud, CNRS, Universit\'e Paris-Saclay,  F-91405, Orsay, France.}

\date{\today}

\begin{abstract}

A study on the spatial organization and velocity fluctuations of non Brownian spherical particles settling at low Reynolds number in a vertical Hele-Shaw cell is reported. The particle volume fraction ranged from 0.005 to 0.05, while the distance between cell plates ranged from 5 to 15 times the particle radius.
Particle tracking revealed that particles were not uniformly distributed in space but assembled in transient settling clusters. The population distribution of these clusters followed an exponential law.
The measured velocity fluctuations are in agreement with that predicted theoretically for spherical clusters, from the balance between the apparent weight and the drag force.
This result suggests that particle clustering, more than a spatial distribution of particles derived from random and independent events, is at the origin of the velocity fluctuations.

\end{abstract}

\maketitle

\section{Introduction}
\label{sec:intro}
%
Particulate flows are of importance in many industrial and environmental applications, and subsequently are subject of research nowadays. Though an apparently simple problem, the settling at low Reynolds number of mono-disperse macroscopic solid particles in a Newtonian fluid is not completely understood. Due to the long range nature of hydrodynamics interactions, the velocity disturbance caused by the motion of a particle decays as slowly as $1/r$ (with $r$ the distance from the particle center).
In the case of the simultaneous settling of several particles, the resulting many-body interactions lead to complex trajectories. 

Indeed, in absence of inertia and in an unconfined Newtonian fluid, one isolated particle settles at the Stokes velocity $U_\mathrm{S}=\frac{2}{9}\frac{{\rho _p-\rho _f}}{\eta }\,g\,a^2$, where $\rho _p$, $\rho _f$, $\eta$, $g$ and $a$ are respectively the particles density, the fluid density, the fluid viscosity, the acceleration due to gravity and the radius of the particles. For a suspension of spheres of volume fraction $\phi$, randomly and independently dispersed in a Newtonian fluid, Batchelor\cite{batchelor72a} calculated a correction to the first order in $\phi$, with average settling velocity $V_\mathrm{sed}  = U_\mathrm{S}\left(1-6.55\phi \right)$. However, for confined suspensions of volume fraction larger than a few percents, there is no theoretical model available, and $V_\mathrm{sed}$ is often described using the empirical correlation\cite{richardson54} $V_\mathrm{sed} = U_\mathrm{S}\, f(\phi)$, where $f(\phi) = \left(1-\phi\right)^n$ ($n\in \left[2, 5.5\right]$, depending of the Reynolds number) is the hindrance function that exists due to the presence of a bottom boundary, and also to the hydrodynamic interactions among particles. 

Due to these hydrodynamic interactions, the settling velocity is constant only in average, and it fluctuates both spatially and temporally.
The standard deviation of the measured particle velocities $\Delta V$  increases with the particle volume fraction up to $\phi \simeq 0.4$ before decreasing due to steric effects.\cite{nicolai95, snabre08a, guazzelli11}  For $\phi \lesssim 0.05$ most of the experimental studies\cite{segre97,snabre08a, guazzelli11} reported $\Delta V \propto V_\mathrm{sed} \, \phi^{1/3}$. These velocity fluctuations are attributed to the permanent evolution of the suspension microstructure: the local volume fraction of the suspension is higher in some regions, and, in those, particles settle faster than the average settling velocity, which in turn, due to the hydrodynamic interactions, modifies the microstructure of the suspension.
Assuming a uniform random spatial distribution of the particles, numerical and theoretical studies\cite{caflish85, hinch88,ladd96} predicted an unrestricted increase of the standard deviation of the velocity fluctuations $\Delta V$ with the vessel size, while Koch and Shaqfeh\cite{koch91} found a single particle spatial distribution that prevents the divergence of $\Delta V$. 
However, velocity fluctuations measured experimentally did not diverge with the size of the vessel. Although, at early times, large-scale fluctuations of size comparable to that of the vessel width were observed, PIV measurements spanning that dimension of the vessel showed that these fluctuations are transient. In the steady-state regime, which is achieved 500 Stokes times $t_{S} = a/U_{S}$ after the beginning of the sedimentation\cite{snabre08a}, the spatial scale of the fluctuation is $l_c \simeq 20\, a\, \phi^{-1/3}$. In the same way, the spatial particle occupancy distribution was found to follow a Poisson law \cite{lei01} at early times, but deviates from a Poisson distribution in the steady-state regime.\cite{lei01, bergougnoux09}
For confined suspensions, the size of the vessel has an influence $\Delta V$ for vessel widths\cite{segre97} $W\leq l_c$ with $\Delta V \propto \phi^{1/2}$ and for vessel thicknesses\cite{brenner99, bernard-michel02} $L \leq 2 \, a \, \phi^{-1/3}$, $\Delta V$ scales as $L^{1/3}$, according to a numerical study.\cite{kuusela04}
Finally, for suspensions confined in capillary tubes, a recent study\cite{heitkam13} reported an average settling velocity larger than $U_\mathrm{S}$. 

In this paper, we investigate how the spatial distribution of the particles affects the velocity fluctuations. Instead of analyzing particle occupancy in a fixed size window, \cite{lei01, bergougnoux09} we studied the way individual particles assemble in groups  (or "clusters"), which may contribute to an increase of the local density, which in turn should impact velocity fluctuations.
The assembly of particles in clusters was characterized by studying the cluster population distribution, i.e. the probability density function of observing a cluster of $N$ particles, and of its statistical moments (average, variance), as a function of the volume fraction $\phi$ and of the ratio $L/a$ between the cell thickness $L$ and the radius of the particles $a$. 
The paper is organized as follows: the experimental setup, methodology and data processing are described in section \ref{sec:expsetup}. Results are presented in section \ref{sec:results}, first describing the statistical properties of the cluster population \ref{sec:cluster}, then analyzing how these properties influence the velocity fluctuations \ref{sec:velfluc}. Conclusions are discussed in section \ref{sec:conclusions}.

 \section{Experimental Set-up}
\label{sec:expsetup}

\begin{figure}[htbp]
\includegraphics[width=0.50 \linewidth]{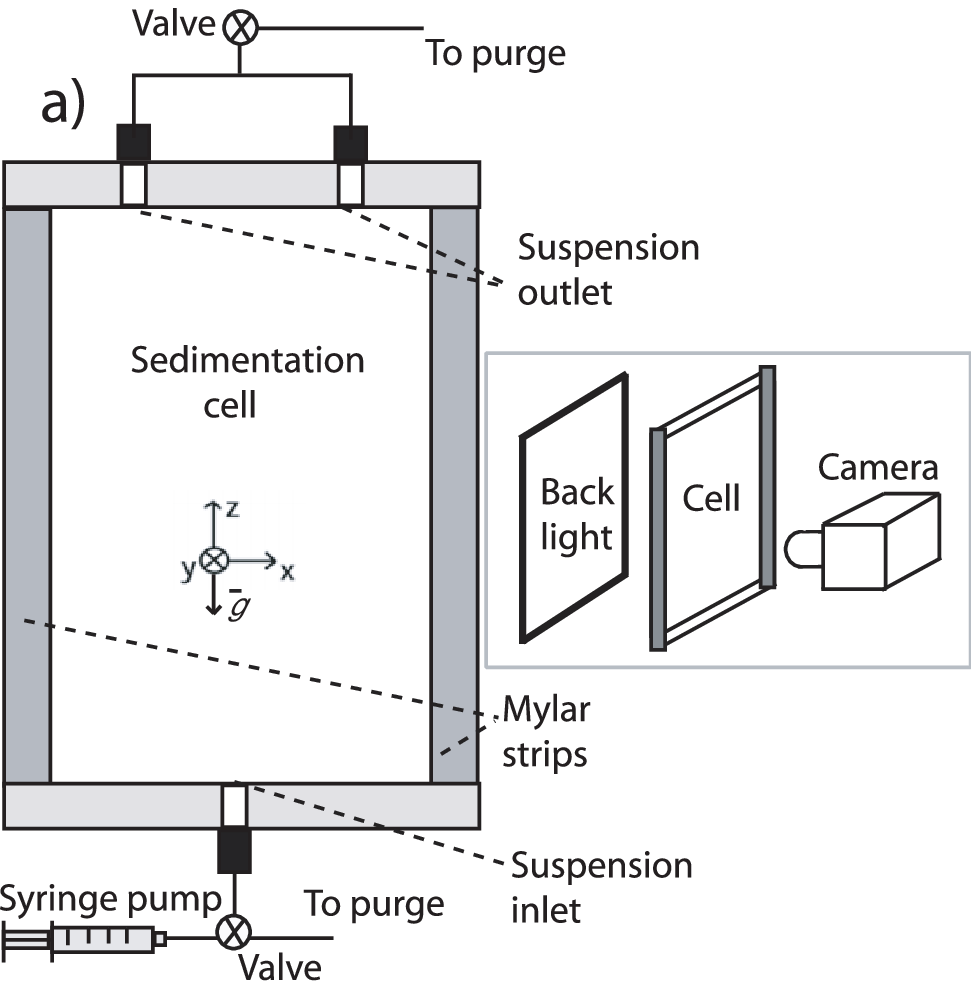} \includegraphics[width=0.45 \linewidth]{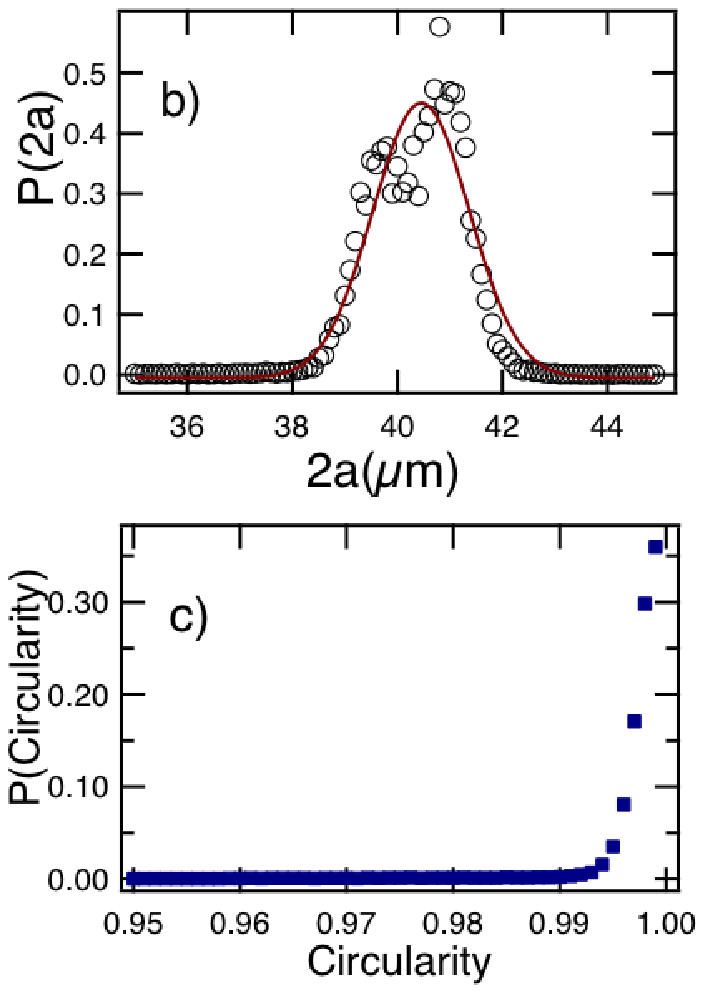}
\caption{a): Scheme of the experimental device and setup. b) and c): Diameter and circularity characterizations of the polystyrene particles used in the experiments. Solid line in b) is a gaussian fit.
}
\label{exp_setup}
\end{figure}

The Hele-Shaw cell (Fig. \ref{exp_setup}.a), set vertically, consisted of two $1$ cm-thick parallel glass plates of $20 \times 15\, \mathrm{cm}$, separated by two mylar spacers of  $20\times 1.5\, \mathrm{cm}$, located along the two vertical sides of the cell.
Mylar thicknesses $L = 100\pm 1,   180\pm 1, \, 250\pm1 \, \mathrm{and}\, 300\pm1 \, \mu\mathrm{m}$ have been used to provide separation between plates $L/a$ = 5, 9, 12.5 and 15. 
 Two perforated parallelepiped plexiglass pieces, with one and two milled holes respectively, were glued to the bottom and top sides of the cell. 
 The bottom hole was connected to the injection syringe, while the top holes were connected to the purge reservoir for drainage. To avoid the presence of microscopic air bubbles after the cell filling, all cells were saturated with $CO_2$ prior to suspension injection.
Finally, to circumvent Boycott effects,\cite{boycott20} great attention has been paid to the verticality of the cell, which was controlled with an uncertainty of $0.3^\circ$. 
Suspensions of volume fraction $\phi \in \left[0.005, 0.05\right]$ were prepared by adding spherical polystyrene particles, of density $\rho_p = 1.05\, \mathrm{g\, cm^{-3}}$ and average radius $a = 20\, \mu\mathrm{m}$, to distilled water, of density $\rho_f = 0.998 \, \mathrm{g\, cm^{-3}}$ at $20^\circ \mathrm{C}$. A small amount of SDS surfactant has been added to the mixture to reduce surface tension in the solid-liquid interface. Characterizations of the diameter and sphericity of the particles were performed using a Morphology G3 equipment (from Malvern Instrument), and are displayed on Fig.~\ref{exp_setup}.b and c. As one can see, the standard deviation is approximately $1\, \mu \mathrm{m}$ for an average diameter $2\, a = 40.5\, \mu\mathrm{m}$, and less than $1\%$ of the particles have a circularity (ratio of the two axes of the ellipse which best fits the perimeter of the particles) bellow $0.99$.   
Suspensions were stirred and then transferred to the injection syringe that was held always vertical to minimize deposition of particles. Then, the suspension was injected into the cell. Finally, once the suspension saturated the cell, valves were closed and the suspension settles freely. This procedure took approximatively 5$s$ which corresponds to 10 Stokes time $t_S = a/U_\mathrm{S}$, largely bellow the duration of the transient regime ($\approx 500\, t_{S}$).

\begin{figure}[htbp]
\includegraphics[width=0.40 \linewidth]{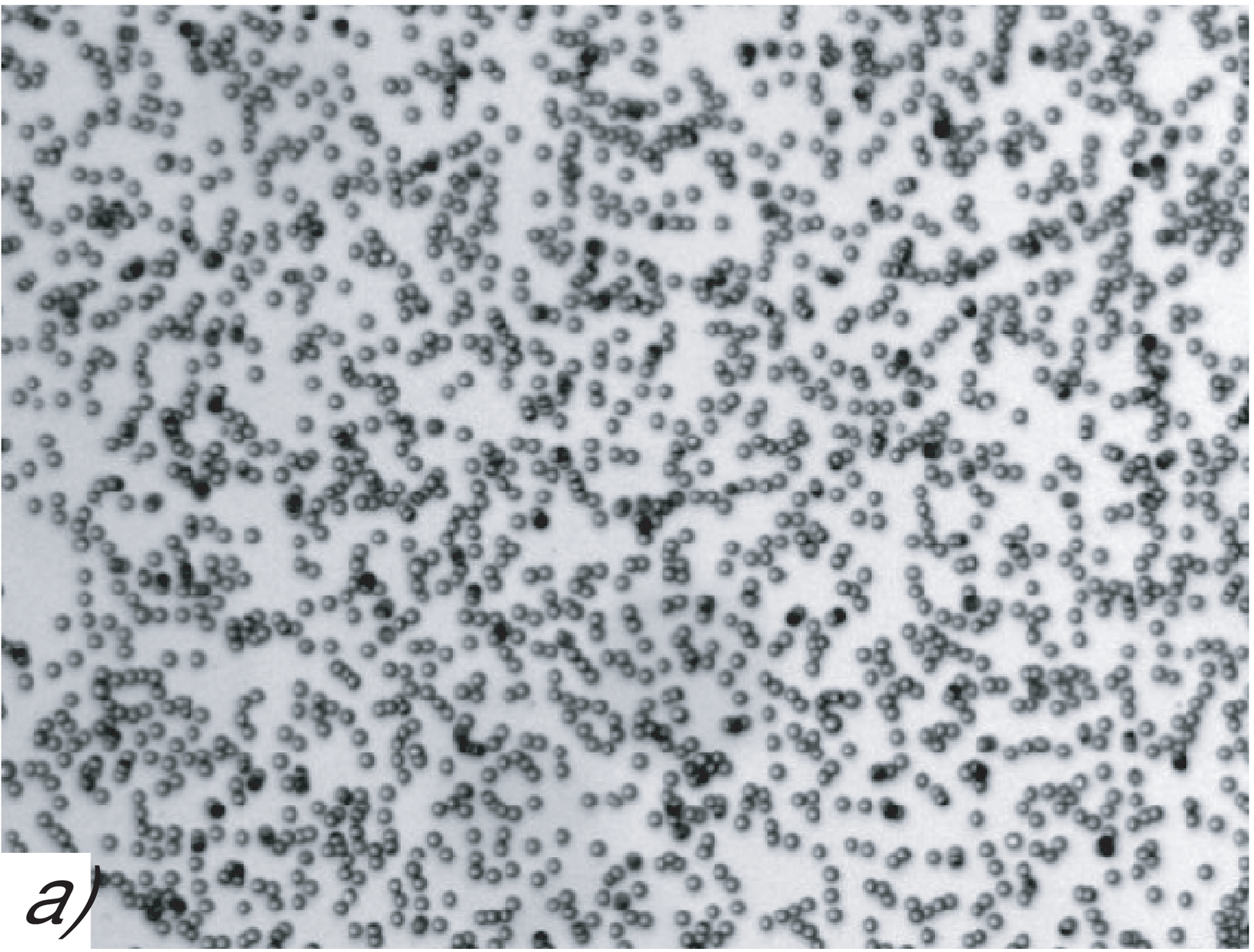}\hskip10pt\includegraphics[width=0.52 \linewidth]{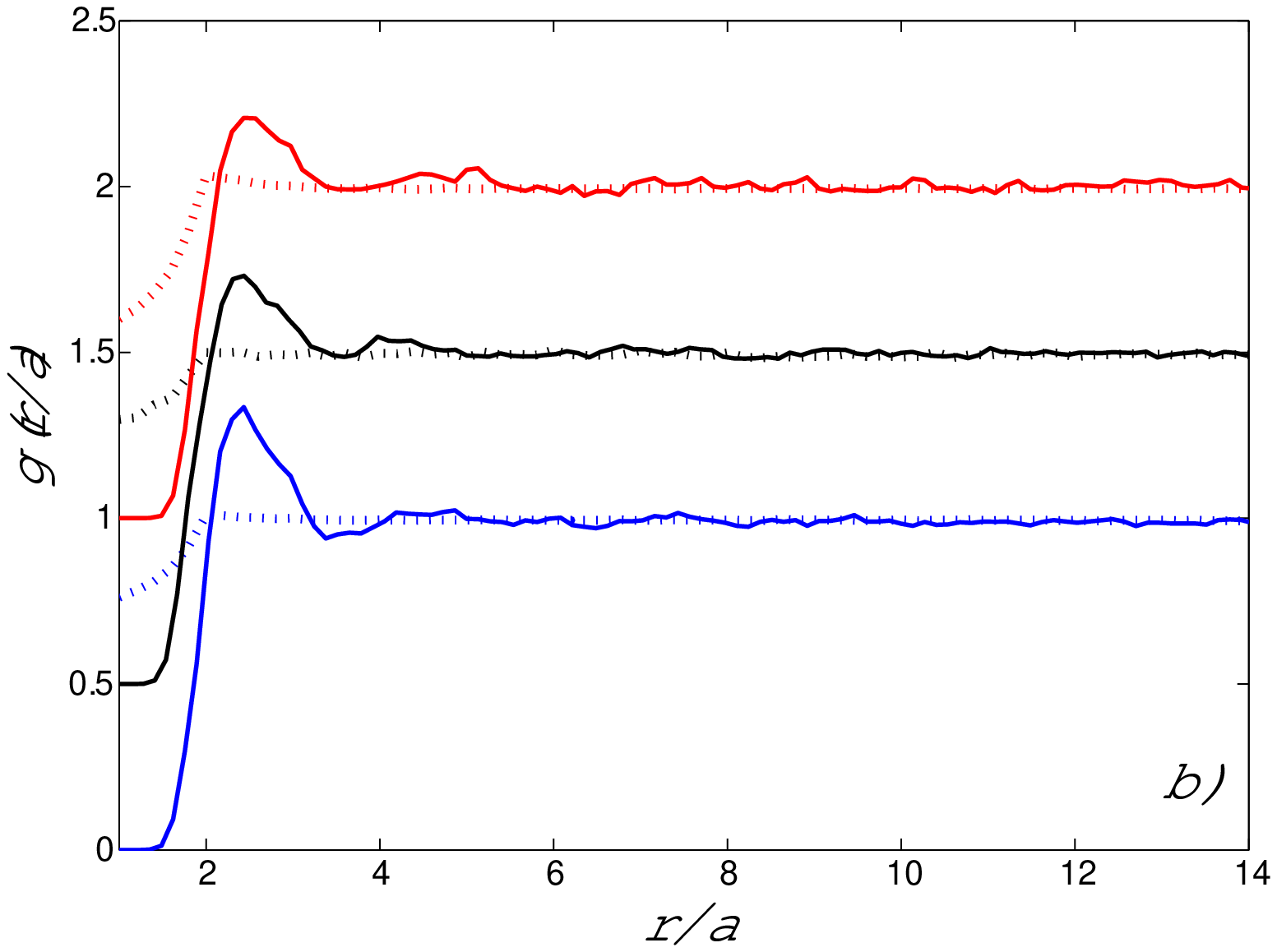}
\caption{a)  A zone of the imaging window as captured by the camera for $L/a$ = 12.5, $\phi$=0.03, $t$= 1800 $s$. b) Radial pair correlation function $g(r/a)$ where $r$ is the distance between particle centers (measured in the $x-z$ plane of Fig.\ref{exp_setup2}(a), as viewed by the camera) and made dimensionless with the particle radius $a$. The data series are displaced vertically by 0.5 for easier visualization. Top solid line: $L/a$ = 15, $\phi$ =0.03. Mid solid line: $L/a$ = 12.5, $\phi$=0.03. Bottom solid line: $L/a$ = 9, $\phi$=0.05. Dotted lines: Pair correlation function calculated numerically for a random configuration of non-overlapping spheres, situated independently, and following a uniform distribution, for the same values of $L/a$ and $\phi$ than the corresponding solid lines.} 
\label{exp_setup2}
\end{figure}

The motion of the particles was captured using a 8-bit CCD camera, located $20$ cm from the cell, with its optical axis perpendicular to the cell plates. The imaging window, situated in the middle of the Hele-Shaw cell, was 2.7 $mm$ width by 3.6 $mm$ high.
The depth-of-field allowed one to visualize particles all across the cell thickness ({\it e.g.} Fig.\ref{exp_setup2}.a).
The positions of the particle centers were detected using a Hough transformation with an uncertainty of less than one pixel (or 1/8 particle diameters). However, if two particles are separated by a distance shorter than $1.5a$, the Hough transform technique used is not capable of detecting both of them.  Stacks of 300 images, captured with a time interval $\delta t = 0.2 \mathrm{s}$ between images, were used to obtain particles trajectories, with a minimal total square displacement rule, and their velocities, using a second order scheme. For each experiment, five stacks were recorded with a time interval $\delta T = 600 \mathrm{s}$ between them, to improve statistics. Finally, to avoid any transient effects, the first stack is acquired $600\, t_\mathrm{S}$ ($264\mathrm{s}$) after the beginning of the sedimentation.

\section{Results and discussion}
\label{sec:results}

\subsection{Cluster population analysis}
\label{sec:cluster}

The number of the detected particles $n_{part}$ in the imaging window was approximately constant in time. Within a stack, the standard deviation of $n_{part}$ divided by its average was $\Delta(n_{part})/\langle n_{part} \rangle \lesssim 0.02$, while $\Delta(n_{part})/\langle n_{part} \rangle \lesssim 0.05$ when considering the five acquired stacks. These fluctuations of the number of particles might be considered negligible, and subsequently, the particle volume fraction $\phi$ was approximately constant during the sedimentation.
Despite the steadiness of the average volume fraction $\phi$ at the scale of the imaging window in time, Fig.\ref{exp_setup2}.a evidences that, as already reported,\cite{ladd02, lei01, bergougnoux09} the spatial distribution of the particles might not be homogeneous. 
As a first step to characterize this distribution, the pair correlation function $g(r/a)$ was computed for the particle positions as detected by the camera. This function represents the probability of finding the center of a particle at a dimensionless distance $r/a$ away from a given reference particle. 
Figure~\ref{exp_setup2}.b displays in continuous lines the experimental $g(r/a)$ for different $\phi$ and $L/a$ combinations. For the same $L/a$ and $\phi$ combinations, in dashed lines it is shown the $g(r/a)$ calculated numerically for a random configuration of particles, situated independently, and following a uniform spatial distribution (except for the hard sphere excluded volume).
In all cases,  the $g(r/a)$ for a uniform distribution shows no evident structure.
In contrast, the $g(r/a)$ for the experiments presents a well defined peak near $r/a \simeq 2.2 $, and some second order structure, independently of $\phi$. This reveals an existing microstructure in the settling suspension.
The non-null value of $g(r/a)$ for $r/a<2$, which might suggest the overlapping of particles, is in fact due to the projection of the actual 3D particle configuration in the $(x,z)$ plane, as detected by the camera.
The peak near $r/a \simeq 2.2$ implies that a significant fraction of the particles settle side by side: during the sedimentation, particles were not isolated but assemble into clusters, with their centers likely to be $2.2 a$ away from each other. Within these clusters, the fluid should have roughly the same velocity as the particles. The presence of a peak at $r/a \simeq 2.2$ has already been reported, using MRI techniques, for a macroscopic suspension settling in a large cell.\cite{talini98}. Besides, clusters are also clearly visible on Fig.4 of the study of Bergougnoux and Guazzelli,\cite{bergougnoux09} which shows the location of particle centers during the sedimentation of a suspension of glass spheres with $a = 75 \mu\mathrm{m}$  and  $\phi =0.003$.

Clusters were detected ``neighbour by neighbour", {\it i.e.}, all the particles with centers closer than $r_c/a = 2.2$ from a reference particle were searched recursively. Once all the particles in a given cluster were identified, it was verified that no one was counted more than once. This procedure allowed one to sort all the particles in sets of clusters of $N$ particles. 
 For completeness, isolated particles were considered as a cluster with $N$=1. In the following, we analyse the population distribution $P(N)$ of the clusters as function of $\phi$ and of $L/a$.

It should be noted that the clusters were identified on the acquired images, in which the real 3D particle spatial configuration was projected  in the $x-z$ plane by the camera. This causes the measured distances between particles to be smaller than the real ones. The projection error in such a measurement increases as $L/a$ increases (would be non-existent for $L/a$ = 2 because all particles would lay in the $x-z$ plane with perfect match between real and projected configurations).
This projection error may lead to an overestimation of the number $N$ of the particles in a cluster. 
While this error cannot be calculated directly, because it depends in the actual 3D spatial configuration of the particles, which is unknown, an upper bound for it was estimated as 0.15 (15\% relative error) for $L/a=15$ (the largest $L/a$ ratio in the experiments). The error decreases with decreasing $L/a$. Details of this estimation are provided in section \ref{sec:appendix}.

\begin{figure}[htbp]
\begin{center}
\includegraphics[width=0.45 \linewidth]{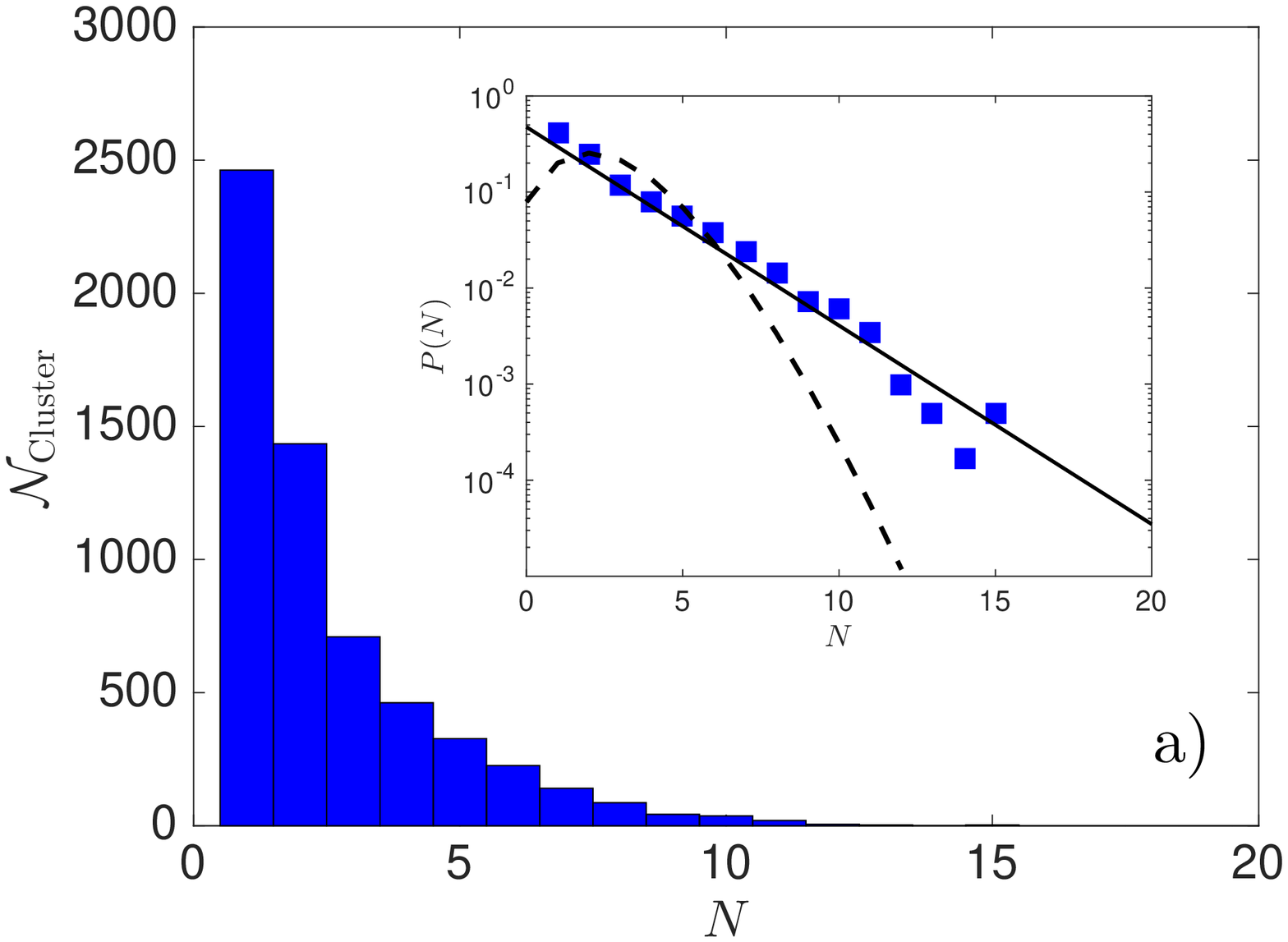} \hskip20pt \includegraphics[width=0.45 \linewidth]{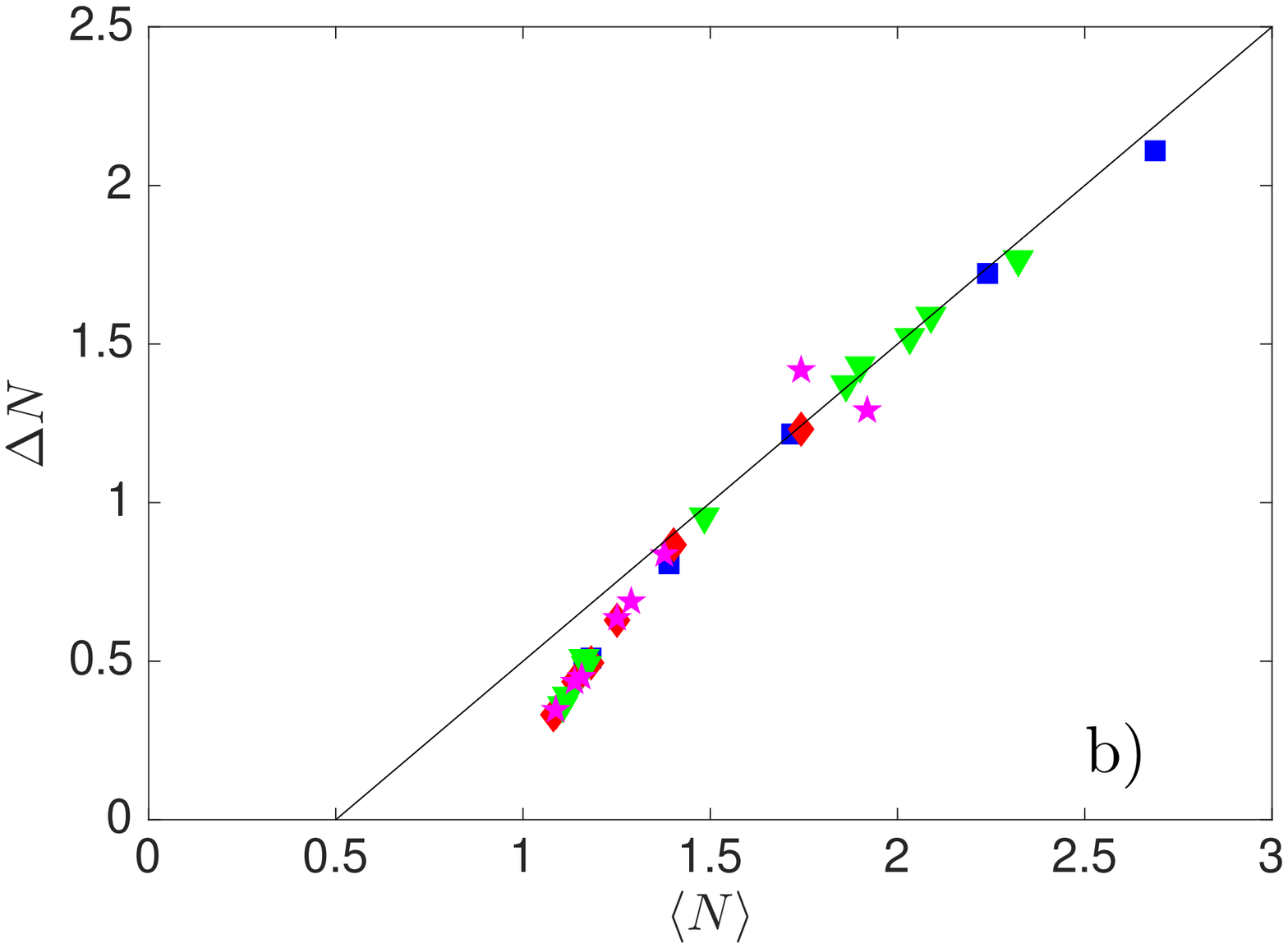}
\end{center}
\caption{a) Histogram showing the number $\mathcal{N}_\mathrm{Cluster}$ of clusters of $N$ particles, for $\phi = 0.053$ and $L/a = 15$. In the inset : Probability density function $P(N)$ of the cluster population $N$. $(\color{blue} \blacksquare \color{black})$ $P(N)$ in semi logarithmic scale and (---) exponential law $P(N) = (1/\langle N \rangle ) \exp\left(-N/\langle N \rangle\right)$, while $(\color{black}{\small - - -} \color{black})$ corresponds to a Poisson distribution $P(N) = \exp\left(-\bn\right) \, \bn^N  / N!$. b) Evolution of the standard deviation $\Delta{N}$ as a function of the average cluster population $\langle N \rangle$. $(\color{blue} \blacksquare \color{black}) \; L/a = 15$, $(\color{green} \blacktriangledown \color{black}) \; L/a = 12.5$, $(\color{magenta} \bigstar \color{black}), L/a = 9$ and $(\color{red} \blacklozenge \color{black})$ \; $L/a = 5$. Solid line (---) corresponds to $\Delta{N} = \bn -0.5$.
}
\label{clus}
\end{figure}

For a given combination of $\phi$ and $L/a$, the number of clusters made of $N$ particles decreases with $N$. This behavior is illustrated on Fig.\ref{clus}.a,  which displays, for  $\phi = 0.053$ and $L/a$ = 15, the number of clusters $\mathcal{N}_\mathrm{Clusters}$ of  $N$ particles as a function of $N$. Once normalized, this distribution corresponds to the probability density function $P(N)$ of cluster population.  $P(N)$ is displayed in the inset of Fig.~\ref{clus}.a as a function of $N$ in a semilog scale. As one can see, $P(N)$ is rather well fitted with an exponential law (solid line) $P(N) = (1/\langle N \rangle ) \exp\left(-N/\langle N \rangle\right)$, where $\bn$ is the average of $P(N))$.

One should note that if the probability that a particle belongs to a given cluster is independent of the population of the latter, this would lead $P(N)$ to follow a Poisson law. The dashed line on the inset of Fig.\ref{clus}.a, which represents this law, shows that this hypothesis is not verified in our experiments.
This is confirmed by the evolution of the standard deviation $\Delta N$ as a function of $\bn$, displayed on Fig.\ref{clus}.b for $L/a = 5,\, 9, \, 12.5$ and $15$ and for $0.005 \leq \phi \leq 0.05$. The data collapse onto a single curve, and for large enough $\bn$ ($\bn \gtrsim 1.5$) $\Delta{N} = \bn -0.5$, in agreement with $P(N)$ following an exponential law.  However, for $\bn \lesssim 1.5$ (roughly isolated particles), we observe a small departure from this linear relation. 

%
%
%

\subsection{Velocity fluctuations}
\label{sec:velfluc}

\begin{figure}[htbp]
\includegraphics[width=0.47\linewidth]{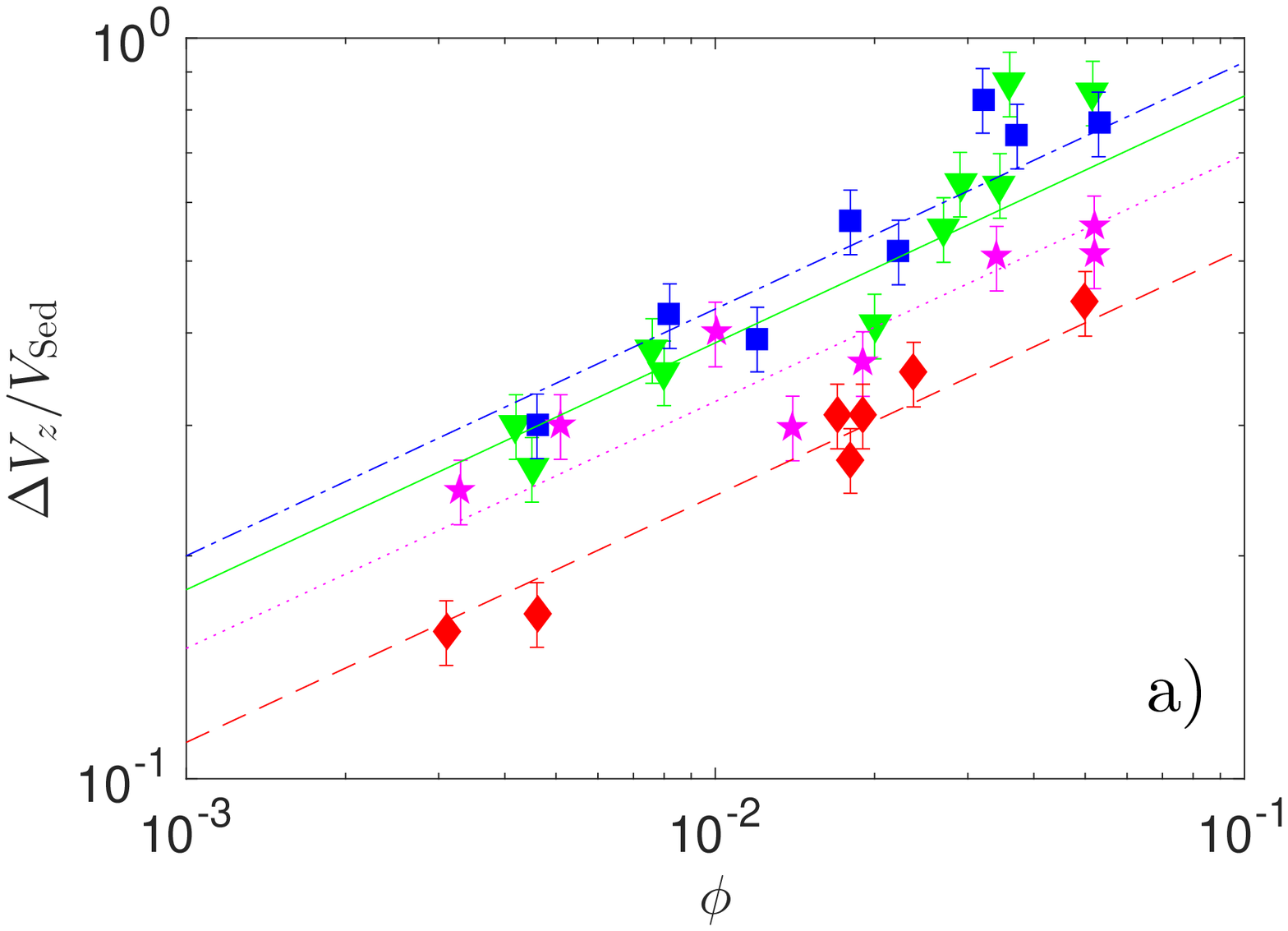} \hskip10pt \includegraphics[width=0.47\linewidth]{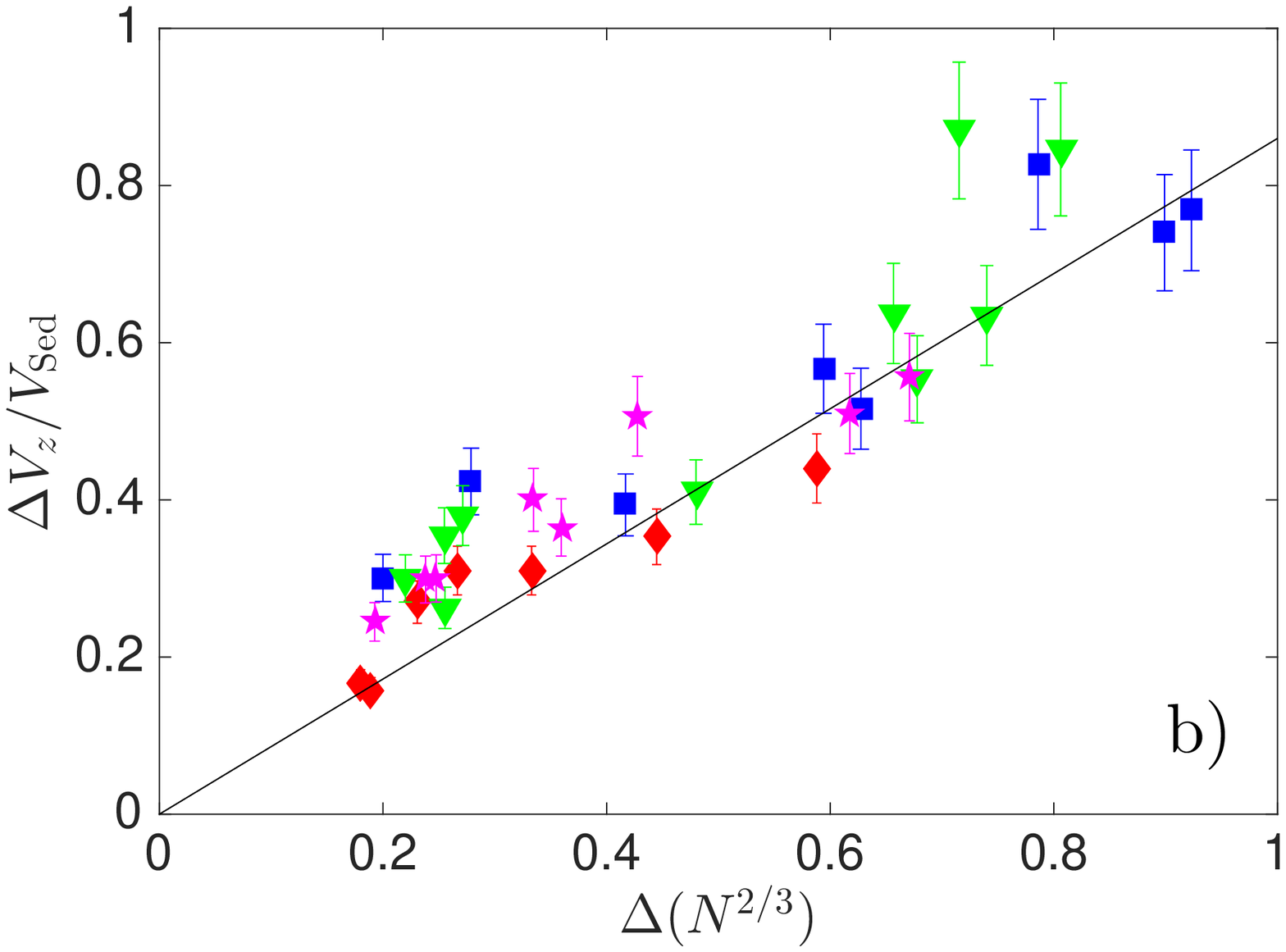}
\caption{Velocity fluctuations $\Delta V/V_{sed}$ as a function of $\phi$ (a), and as a function of the standard deviation of $N^{2/3}$ (b).  $(\color{blue} \blacksquare \color{black}) \; L/a = 15$, $(\color{green} \blacktriangledown \color{black}) \; L/a = 12.5, (\color{magenta} \bigstar \color{black}) \; L/a = 9$ and  $(\color{red} \blacklozenge\color{black})$ \; $L/a = 5$.  
$(\color{blue} \cdot - \color{black})$, $(\color{green} \text{---}  \color{black})$, $(\color{magenta} \cdot\cdot\cdot \color{black})$ and $(\color{red} -- \color{black})$ correspond respectively to best fits over $(\color{blue} \blacksquare \color{black})$, $(\color{green} \blacktriangledown \color{black})$, $(\color{magenta} \bigstar \color{black})$ and $(\color{red} \blacklozenge \color{black})$ with the expression $\Delta V/V_{sed} = \alpha \times\phi^{1/3}$. These fits yield $\alpha = 1, 1.5, 1.8$ and $2$.
The solid line in (b) is $\Delta V/V_{sed} = 0.86\,\Delta (N^{2/3})$.}
\label{deltav}
\end{figure}

The velocity fluctuations were characterized by the standard deviation of the particle velocities normalized by their average. These magnitudes were calculated over all the particles in the last 299 images of each stack. The velocity fluctuations obtained in this way for the five different stacks captured in each experiment were in agreement within a $10\%$ variation and, in the following, $\Delta V/V_\mathrm{sed}$ corresponds to an average over the five stacks. The relatively small variation of $\Delta V/V_\mathrm{sed}$ over the different stacks confirms that, in the steady-state regime,\cite{lei01, snabre08a} the velocity fluctuations have no significant evolution during the sedimentation. 
Figure~\ref{deltav}.a displays $\Delta V/V_\mathrm{sed}$ in logarithmic scale, as a function of $\phi$ for the four values of $L/a$ studied. For all $L/a$, $\Delta V_z/V_{sed} \simeq \alpha \phi^{1/3}$, in agreement with previous studies\cite{guazzelli11} and a theoretical prediction that accounts for the presence of confining walls.\cite{brenner99} Indeed, for $L/a \geq 9$ the volume fraction was large enough, $\phi \geq \phi^*$ ($\phi^*\lesssim 0.008, \, 0.004$ and  $0.0023$ for $L/a =9; 12.5$ and $15$ respectively) so that particles interact with each other more than with the walls. It has to be noted that this trend exists even for $L/a = 5$ while $\phi < \phi^* \approx 0.06$. Moreover, best fits of the evolution of $\Delta V/V_\mathrm{sed}$ with $\phi^{1/3}$ leads to $\alpha = 1,\, 1.5,\, 1.8$ and $2$ for $L/a =5,\, 9,\, 12.5$ and $15$ respectively, in agreement with $\alpha \propto \left(L/a\right)^{1/3}$ reported in a numerical study.\cite{kuusela04}%

\noindent To study the connection between velocity fluctuations and the cluster population distribution, we followed Caflish,\cite{caflish85} Hinch,\cite{hinch88} and Rouyer {\it et al.}\cite{rouyer00} who related the velocity fluctuations to the statistical fluctuations of the spatial distribution of the particles by considering a "blob", {\it i.e.} a given region of space with an excess of particles. Balancing the apparent weight of the blob with its Stokes drag, they calculated its excess of velocity. Using the same approach, we consider here a cluster of $N$ particles. The apparent weight of the cluster is : $\vec{P} = N v_p(\rho_p - \rho_f) \vec{g}$ where $\rho_p$ and $\rho_f$ are the densities of the particles and the fluid respectively, while $v_p=4/3 \pi a^3$ is the volume of a particle. 
Assuming spherical clusters of radius $R_c$, the Stokes drag may be written as: 
$\vec{F}  = - 6 \pi \eta\, R_c \,f^{-1}(\phi)\, \vec{V}_c$ where $\vec{V}_c$ is the velocity of the cluster, $\eta$ the fluid viscosity and $f(\phi)$ is the hindering function\cite{richardson54} that takes into account the backflow due to the confinement.		
Then, balancing the drag with the apparent weight yields $\vec{V}_c = f(\phi) \,\left(N v_p \left(\rho_p - \rho_f\right) \vec{g}\right) / \left(6 \pi \eta R_c\right)$. Finally, writing $R_c = \left(N/\phi_m\right)^{1/3}a$, where $\phi_m$ is the effective volume fraction of the cluster, one obtains: 
${V}_c = \phi_m^{1/3} f(\phi) U_\mathrm{Stokes} \, N^{2/3} $ and subsequently, since $\langle V_c \rangle = f(\phi) U_\mathrm{S}$, the standard deviation of the cluster velocities reads:
\begin{equation}
\frac{\Delta V_c}{\langle V_c\rangle} = \phi_m^{1/3} \Delta (N^{2/3}) \label{dv}
\end{equation}
\noindent where $\Delta (N^{2/3})$ is the standard deviation of  $N^{2/3}$.

\noindent Figure~\ref{deltav}.b displays the standard deviation of the vertical velocities $\Delta V_z$ normalized with the average settling velocity $V_\mathrm{sed}$ as a function of $\Delta (N^{2/3})$.
The data collapse fairly well onto a single master curve with a linear trend, for all $\phi$ and $L/a$. The continuous line on Fig.~\ref{deltav}.b has a slope of 0.86 which would correspond to spherical clusters of a random close packing of spheres ($\phi_m^{1/3} \approx 0.86$ for $\phi_m = 0.64$). As one can see, experimental data are slightly above the prediction for spherical clusters.

The fact that velocity fluctuations can be strongly related to inhomogeneities in the particle spatial distribution has been shown theoretically \cite{hinch88}, and some authors \cite{lei01, bergougnoux09} characterized this inhomogeneity in a fixed inspection window. This result extends the validity of previous findings\cite{hinch88, rouyer00, lei01, bergougnoux09} by determining the relation between velocity fluctuations and the population of particle clusters, rather than particle distribution in a fixed inspection window.

\section{Conclusions}
\label{sec:conclusions}
The spatial distribution of particles in a settling suspension has been studied. The pair correlation function of the particle positions have revealed a peak for a distance of 2.2 particle radius between particle centers, which suggested a cutoff length for defining clusters of settling particles. The distribution of the number of particles in the clusters have been found to follow an exponential law. The average and the standard deviation of this distribution increase with the particle volume fraction $\phi$, while the ratio $L/a$ appears to have only weak influence in the range studied.

The measured velocity fluctuations were rather well predicted assuming that particles assemble in spherical clusters.

The discrepancy between the experimental result and the predicted value of 0.86 (Figure~\ref{deltav}.b)  could be related to the fact that the particle diameter distribution (Fig. \ref{exp_setup}.b) presented a small degree of  polidispersity, which might increase the value of $\phi_m$ compared with monodisperse spheres.  However, results by other authors \cite{farr09} indicate that for such a narrow distribution, this increase is likely to be negligible. Another possible explanation is that clusters are not perfectly spherical but more prolate spheroids: A prolate spheroid with its longest axis aligned with the gravity direction would settle faster than one with its longest axis perpendicular to the gravity direction. The resulting fluctuations in the settling velocity, if both axis alignments coexist, would then be larger than for spherical only clusters. To be conclusive on this aspect of the velocity fluctuations, one should calculate the probability density function of the clusters aspect ratio and of their orientation with respect to gravity, which would require a large amount of detected clusters to achieve a good statistical sampling. While such a description is beyond the scope of the present study, it constitutes an interesting motivation for future work.

\section{Acknowledgements}
\label{sec:ack}
The authors would like to thank N. Torres Cabrera for contributing to the development of the tracking code and for preliminary experiments, and L. Oger, D. Salin, F. Rouyer and G. Drazer for fruitful discussions. This research has been supported by PIP 0246 CONICET, ANPCyT PICT-2013-2584, and the LIA-FMF in Physics and Mechanics of Fluids.

\section{Appendix}
\label{sec:appendix}

The measurement error due to projection could not be directly calculated, because it depends in the actual 3D spatial configuration of the particles, which is unknown.
If particles are closer to each other than in the case of a uniform random distribution, forming clusters, as suggested by the peak in the experimental $g(r/a)$, and as it is the thesis of the present work, the error calculated for such a distribution may provide an upper bound for the error in the experimental configurations.

As stated in the manuscript, two particles participate in a cluster if they are less than $2.2a$ away from each other. The projection error can be then quantified by comparing the probability for two particles being less than $2.2a$ away from each other in the 2D projection, with the same probability in the actual 3D particle spatial configuration. As the first probability exceeds the second one, the error increases.

From the $g(r/a)$ curves shown in Figure~\ref{exp_setup2}.b, it can be noted that, in the 2D projection, if two particles are separated by a distance shorter than $1.5a$, the Hough transform technique used is not capable of detecting both of them.
Taking this into account, the first probability reads:

$$P_{2D} =\frac{1}{\pi R^2} \int^{2.2a}_{1.5a} g_{2D}(r) 2\pi r dr$$

while the second one reads :
 
$$P_{3D} =\frac{1}{\frac{4}{3}\pi R^3} \int^{2.2a}_{0} g_{3D}(r) \pi r^2 dr$$

where $g_{2D}$ is the $g(r)$ calculated from the 2D projection of the particle positions as viewed by the camera, and $g_{3D}$ is the $g(r)$ of the actual particle distances in 3D spatial configuration.

The relative error can be written as: $E=2(P_{2D} - P_{3D})/(P_{2D} +P_{3D})$. For $\phi = 0.05$ and $L/a=15$ (the largest $L/a$ ratio in the experiments), this estimation yields $E = 0.15$ or a 15\% relative error. The error $E$ decreases as $L/a$ decreases.

%
%

\end{document}